\documentclass{elsart}

 \usepackage{graphics}
 \usepackage{graphicx}
 \usepackage{epsfig}

\usepackage{amssymb}

\begin{document}

\begin{frontmatter}


 \title{A Study of the Scintillation Induced by
 Alpha Particles and Gamma Rays in Liquid Xenon in an Electric Field}



\author[ImperialCollege]{J. V. Dawson}\ead{jaime.dawson@imperial.ac.uk},
\author[ImperialCollege]{A. S. Howard},
\author[ITEP]{D. Akimov},
\author[ImperialCollege]{H. Araujo},
\author[ImperialCollege]{A. Bewick},
\author[ImperialCollege]{D. C. R. Davidge}, 
\author[ImperialCollege]{W. G. Jones}, 
\author[ImperialCollege]{M. Joshi}, 
\author[ImperialCollege,ITEP]{V. N. Lebedenko}, 
\author[ImperialCollege]{I. Liubarsky},
\author[ImperialCollege]{J. J. Quenby},
\author[ImperialCollege]{G. Rochester},
\author[ImperialCollege]{D. Shaul},
\author[ImperialCollege]{T. J. Sumner},
\author[ImperialCollege]{R. J. Walker}

\address[ImperialCollege]{Department of Physics, Blackett
Laboratory, Imperial College London, \\SW7 2BW, UK}
\address[ITEP]{Institute of Theoretical and Experimental Physics, Moscow, RU}

\begin{abstract}
Scintillation produced in liquid xenon by alpha
particles and gamma rays has been studied as a function of applied electric
field.  For back scattered gamma rays with energy of about 200 keV, the
number of scintillation photons was found to decrease by 64$\pm$2\% with increasing field strength.  Consequently, the pulse shape discrimination power between alpha particles and gamma
rays is found to reduce with increasing field, but remaining non-zero at higher fields.  
\end{abstract}

\end{frontmatter}

\section{Introduction}
\label{Introduction}

There is much recent interest in xenon as a target for Weakly Interacting
Massive Particle (WIMP) detectors.  It has a high atomic mass of 131
amu, good for maximising the energy deposition from a WIMP collision.  This is due to a close match between the nuclear mass and the most favoured WIMP mass according to SUSY models.  Xenon is a
scintillator, giving $\sim$~52~$\times$~$10^{3}$ photons/MeV for recoiling
electrons in our energy range of interest~\cite{Doke2002,Barabanov1987}.  The emitted photons are VUV, from an emission band centred at 176.7~nm \cite{Schwentner1985}, which can be detected with
photomultiplier tubes.

The observed scintillation pulse shape has a fast rise ($\sim$ few ns)
followed by an exponential-like decay.  The decay time is
indicative of the incident particle, allowing discrimination between
gamma rays,alpha particles and nuclear recoils~\cite{Kubota1979,Akimov2002}.
This property is exploited by ZEPLIN I \cite{Hart2002}, the Boulby Dark Matter
Collaboration's liquid xenon dark matter detector.  The scintillation
pulse shape is used to differentiate between background gamma rays
and recoiling nuclei produced by scattering WIMPs.  

The next generation of xenon dark matter detectors are two-phase
systems; ZEPLINs II~\cite{Cline2000} and III~\cite{Dawson2003}.  Such detectors
measure both scintillation and ionisation produced by interacting
particles separately.  They comprise a liquid xenon target and gaseous
xenon layer.  An applied electric field removes electrons from the
primary interaction site and extracts them into the gas phase.  Whilst
crossing the gas layer, the electrons have sufficient kinetic energy
to excite the xenon gas atoms producing electroluminescence.  VUV
signals are produced by both the primary scintillation and the subsequent secondary electroluminesence.  In a uniform electric field, the
electroluminesence signal is directly proportional to the number of
electrons removed from the interaction site.  The ratio of primary to
secondary signal sizes is used as a discrimination tool for the
two-phase system.

This work presents a study of the scintillation
signal, produced by alpha particles from an \nuc{241}{Am} source and gamma
rays from a \nuc{60}{Co} source, as a function of applied electric field.

The VUV scintillation signal from liquid xenon is produced by the
de-excitation to ground of the two lowest excited molecular states;
$^{1}\Sigma^{+}_{u}$ and $^{3}\Sigma^{+}_{u}$ .  These states can be
produced initially through excitation, or via recombination~\cite{Kubota1979}.  The lifetimes of these states are found to be 4.3~$\pm$~0.6~ns and 22.0~$\pm$~2.0~ns for alpha particles~\cite{Hitachi1983} and  2.2~$\pm$~0.3~ns and 27~$\pm$~1~ns for
recoiling electrons~\cite{Kubota1979}.  In addition, there is a long
recombination time of $\sim$~15~ns for recoiling
electrons~\cite{Kubota1979}.  These effects are convoluted into a
pulse shape that can be approximated as a single exponential decay.
This is particulary true with low numbers of photons where the Poisson
width blurs any distinction between the respective components.  In
this work, the overall decay profile for recoiling electrons has a
measured mean time constant of 48.4~$\pm$~0.2~ns compared to 15.0~$\pm$~0.2~ns for alpha particles.

Under an electric field, the gamma ray induced (or recoiling electron) pulse
shape and area changes.  This is due to the removal of electrons from
the interaction site, suppressing the recombination component of the
scintillation~\cite{Conti2003}.  Understanding the behaviour of gamma
ray signals under an electric field is vital to two-phase xenon
detectors in order to calibrate the energy scale, and estimate background
contributions etc.  To predict the primary and secondary signal sizes the
number of electrons removed from a gamma ray interaction as a function
of energy deposit and electric field strength is needed.  For charge
measurements, a linear relationship is observed between the inverse of
the charge extraction and the inverse of the electric field~\cite{Jaffe1913,Kubota1978}.  Using this functional form, the
proportion of scintillation lost (or charge gained) can be found for
an infinite electric field.  In the work presented here,
recombination is found to be responsible for 64~$\pm$~2~\% of the total
scintillation signal for gamma rays.  Normalising by this quantity allows direct comparison between
previous charge measurements.  The proportion of the
recombination component removed by the electric field, $N_{e}(\infty)$, is
defined as 1 at an infinite electric field.  The behaviour of
the quantity ($N_{e}$) versus electric field ($E$) is dependent on
the deposited gamma ray energy ($E_{\gamma}$).  Charge measurement
data of Voronova~\cite{Voronova1989}, with gamma ray energies of 15.3,
17.3, 21.4 and 662~keV, show this dependence.  It is possible to
parametrise their data as follows \cite{Davidge2003}:  

\begin{equation}
\frac{1}{N_{e}} = \left(\frac{30.07}{E_{\gamma}} + 0.40 \right)
\frac{1}{E} + 1, \label{eqn:davidgeparameterisation}
\end{equation}
where $E_{\gamma}$ is the energy of the incident gamma ray in keV and $E$
is the electric field strength in kV/cm.

The gamma ray recombination component is suppressed by an applied
electric field, causing the overall decay time of the scintillation
pulse to reduce.  The alpha particles pulse shapes show
no discernable change with increasing the electric field strength up to
3.7~kV/cm.  The gamma ray decay profile changes from a time
constant of 48.4~$\pm$~0.2~ns with no electric field to
23.7~$\pm$~0.2~ns at 3.7~kV/cm.  This change can be used in a
diagnostic capacity in two phase detectors since gamma rays
interacting in low electric field regions will present longer decay
profiles.  Such regions can be problematic in a two phase detector as
interacting gamma rays may not suffer sufficient charge extraction to
produce electroluminesence, or smaller electroluminesence signals may
result.  Whilst the gamma ray scintillation decay time is reduced by
an electric field, under low fields it is still possible to
discriminate between gamma rays and nuclear recoils using Pulse Shape
Discrimination.

\section{Experimental Setup}
\label{Experimental_Setup}

\begin{figure} 
\centering{ 
\scalebox{0.5}{\includegraphics{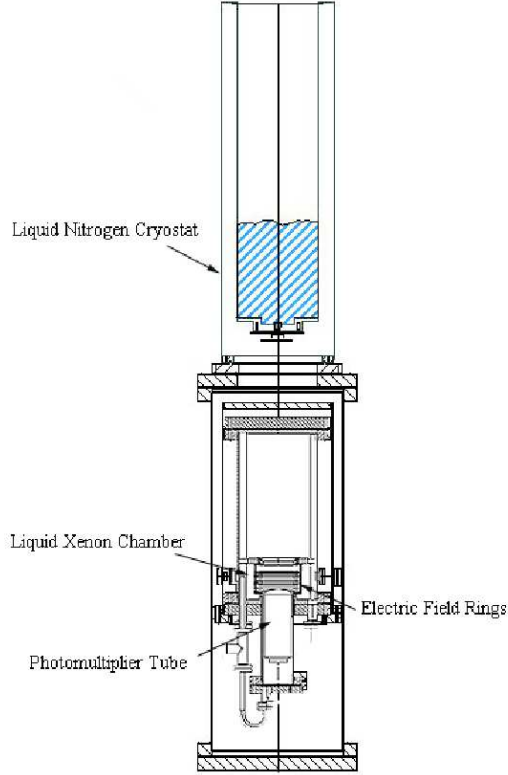}}
\caption[\small Schematic of Detector.]{\small Schematic of the detector
  showing liquid nitrogen cryostat and the liquid xenon chamber
  containing the photomultiplier tube and electric field ring
  structure.  \label{fig:prototype}}}
\end{figure}

\begin{figure} 
\centering{ 
\scalebox{0.5}{\includegraphics{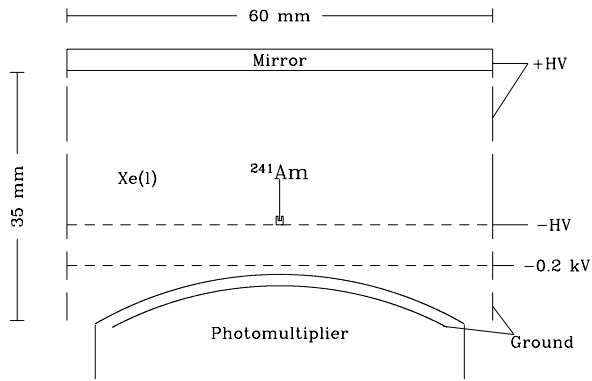}}
\caption[\small Detector Schematic ]{\small Schematic of the internal structure,
showing the electric field rings, top electrode (mirror), grids and face of the
photomultiplier tube.  The \nuc{241}{Am} source and holder are
shown.  The voltage loads are applied as shown on the right-hand side
of the diagram.  \label{fig:prototypediag}}}
\end{figure}

A prototype two phase liquid xenon dark matter detector was used for
these studies, Figure \ref{fig:prototype}.  Inside
the chamber is a photomultiplier tube immersed in liquid xenon, field
shaping rings, two grids, an aluminium mirror (electrode) and a fixed
\nuc{241}{Am} internal source.  The total xenon mass is 4.5 kg.  The sensitive
liquid xenon volume is defined by the field rings and for this
experiment fills the structure to the mirror (top electrode).  The
field rings have an internal radius of 3 cm.  Figure
\ref{fig:prototypediag} is a diagram of the sensitive volume of the
xenon chamber, showing the electric field rings, top electrode
(mirror), high voltage grids and photocathode of photomultiplier tube.
The \nuc{241}{Am} source is fixed to the upper high voltage grid and
sits upward facing inside a lead holder.  Voltages are applied to the
upper high voltage grid (-ve) and mirror (+ve), providing the drift
field.  The lower grid is kept at a low negative potential (-~0.2~kV)
primarily to shield the photocathode from the high voltage grid.  The photomultiplier tube is
quartz windowed (ETL 9829B) with a photocathode quantum efficiency of
23~\% at 175 nm \cite{ETL2003}.

%

\subsection{Xenon Purity}
\label{Xenon_Purity}

In order to drift electrons for centimetres without loss through xenon
requires high purity (parts per trillion): in particular, free of
electronegative contaminants such as O$_{2}$, N$_{2}$O, CO$_{2}$ and
H$_{2}$O.  The
scintillation is also affected by the presence of impurities
(shortening the total attenuation length) but to a lesser extent
(parts per billion).

Purification apparatus and procedures have been developed
to remove contaminants~\cite{Howard2001}.  All components of the inner
chamber were constructed from stainless steel, copper, quartz or high
purity ceramic, and designed for ultra-high vacuum use with residual
pumped pressures below 10$^{-8}$~mbar.  The xenon was delivered to an Oxisorb cartridge~\cite{Oxisorb2003} which removes
O$_{2}$ and H$_{2}$O chemically.  Xenon was distilled into the
detector from a temperature controlled external bottle through a
0.5~$\mu$m dust filter.  This
distillation procedure separates xenon from other trace impurities
left solidified behind, particularly CO$_{2}$, CH$_{4}$ and H$_{2}$O.

The xenon is kept cold via liquid nitrogen delivered to a cryostat
directly above the main chamber.  The chamber temperature is monitored
at various locations, and with the use of heaters and a restrictive
coupling the liquid temperature is actively maintained at -97$^{\circ}$C.

The purity of the xenon is assessed via the measurement of the free
electron lifetime.  To do this, potential differences are
applied between the high voltage grid and top electrode causing free
electrons released by background gamma and cosmic ray interactions and
alphas from the internal \nuc{241}{Am} source to drift towards the
mirror.  The electron drift is quantified with a calibrated
charge sensitive amplifier (Amptek A250) connected to the top electrode
(mirror).  The outputed voltage is recorded as a function of time,
triggered by the primary scintillation signal.  The longest drift
times are found to be in excess of 60~$\mu$s~ across a 21.75~mm drift
path~\cite{Howard2001}, indicating a very high purity, without any
measurable loss of charge even at fields below 10 V/cm.


\subsection{Applied Electric Field}
\label{Applied_Electric_Field}

The electric field is produced by applying negative voltages to the
upper high voltage grid and positive voltages to the top electrode.
The total potential difference between the upper high voltage grid and
top electrode ranged from 0 to 8~kV.  The corresponding electric field
distribution was modelled using the finite element analysis code,
ANSYS~\cite{ANSYS2003}.

\section{Calibration}
\label{Calibration}

\begin{figure}
\centering{
\scalebox{0.4}{\includegraphics{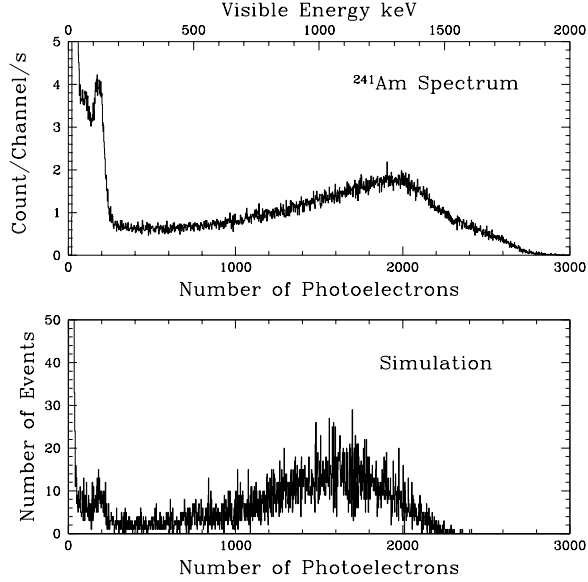}}
\caption[\small Pulse Height Spectrum of Internal \nuc{241}{Am}
Source. ]{\small Pulse Height spectrum of internal \nuc{241}{Am}
source (upper plot) and GEANT~4 simulated spectrum (lower plot).  The alpha distribution is broad and peaks at $\sim$1900 photoelectrons.  The 60~keV gamma peak is found at $\sim$~180~photoelectrons.    \label{fig:am241spectrum}}
}
\end{figure}

The calibration was made with no applied electric field.  Initially the scintillation spectrum from the internal source was measured.  The
  \nuc{241}{Am} source is coated, producing alpha particles
  with a broad energy distribution peaking at 3.9 MeV.  As the source
  is mounted in a lead source holder, the scintillation light produced
  by the alpha particles has no direct path to the phototube (maximum
  shadowing).  Only light reflected by the mirror is seen.  The peak
  of the alpha distribution contains $\sim$1900 photoelectrons as
  shown in Figure \ref{fig:am241spectrum} (upper spectrum).  The
  resulting spectrum produced by a GEANT~4~\cite{Geant42003}
  simulation of the full detector and \nuc{241}{Am} source is also
  shown (lower spectrum).

\begin{figure}
\centering{ 
\scalebox{0.4}{\includegraphics{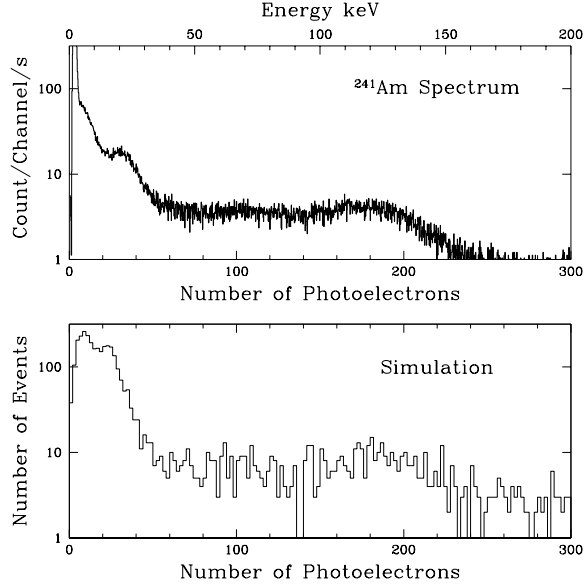}}
\caption[\small Pulse Height Spectrum of Internal \nuc{241}{Am}
Source. ]{\small Pulse Height spectrum of internal \nuc{241}{Am}
source (upper plot) and GEANT~4 simulation spectrum (lower plot).  The
60~keV gamma rays from the $^{241}$Am source result in 3~peaks.  The uppermost peak is from gamma rays penetrating the sides of the source holder giving direct light to the phototube.  The lower peak (at $\sim$20~photoelectrons) is from 60~keV gammas maximally shadowed by the source.  The intermediary structure
is from partial shadowing of the direct light and multiple interactions.
\label{fig:loweam241spectrum}}
}
\end{figure}

The \nuc{241}{Am} source produces 60~keV gamma and X-rays, resulting in a
peak at $\sim$~180 photoelectrons.  The low energy spectrum is shown in
Figure \ref{fig:loweam241spectrum} with the corresponding simulated
spectrum.  The 60~keV gamma rays are seen as 3~peaks; the uppermost
($\sim$~180~photoelectrons) giving direct light to the phototube from
gamma rays passing sideways through the lead source holder, the
lowermost ($\sim$~20~photoelectrons) from gamma rays emitted upwards
being maximally shadowed by the source holder, and the intermediary
structure resulting from partial shadowing and multiple interactions.  Below these three peaks, lower energy gamma rays from \nuc{241}{Am}
(maximally shadowed) merge to form the rest of the low energy
spectrum.

The upper 60~keV line results from events occuring less than 3~mm from
the \nuc{241}{Am} source, with most interactions occuring at a
distance of 1~mm.  These scintillation signals with no applied
electric field, give a measured yield of 3~photoelectrons per keV.  
These interactions occur in the most efficient
light collection region in the chamber, with the rest of the chamber
being a factor 2.3 $\pm$ 0.3 less efficient.  The spectra were energy
calibrated assuming 1.5~photoelectrons per keV.

\begin{figure}
\centering{ 
\scalebox{0.4}{\includegraphics{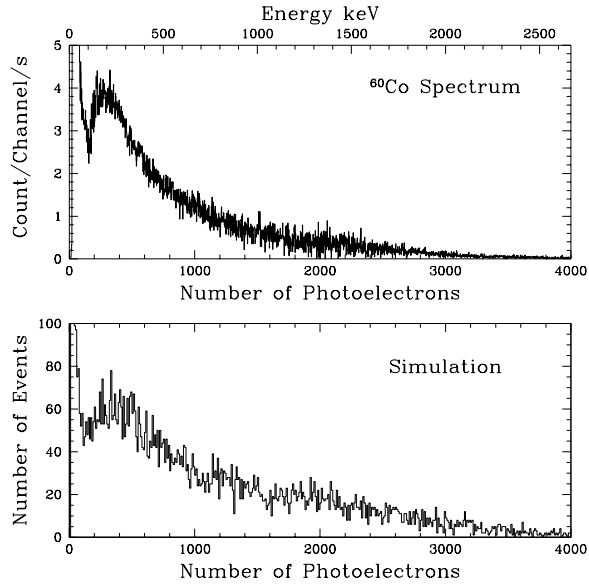}}
\caption[\small Pulse Height Spectrum of External \nuc{60}{Co}
Source.  ]{\small Pulse Height spectrum of external \nuc{60}{Co}
  source (background subtracted).  The lower plot is a GEANT~4
  simulated spectrum.  The broad peak at $\sim$~200~keV is from
  gamma rays scattered at large angles into the sensitive xenon volume, and comprises $\sim$~300~photoelectrons.   \label{fig:co60spectrum}}
}
\end{figure}

Figure \ref{fig:co60spectrum} is a spectrum from an external
\nuc{60}{Co} source after subtraction of the background spectrum from
the internal source (Figure \ref{fig:am241spectrum}).  Due to the
amount of liquid xenon surrounding the sensitive region, the spectrum
displays only one clear feature: a broad peak from gamma rays
scattering at large angles in the external xenon volume and depositing
$\sim$~212~keV in the internal xenon volume.  This feature was replicated in
the simulation (lower plot).

\section{Pulse Height Spectra Analysis}
\label{pulse_height_spectra}

\begin{figure}
\centering{ 
\scalebox{0.4}{\includegraphics{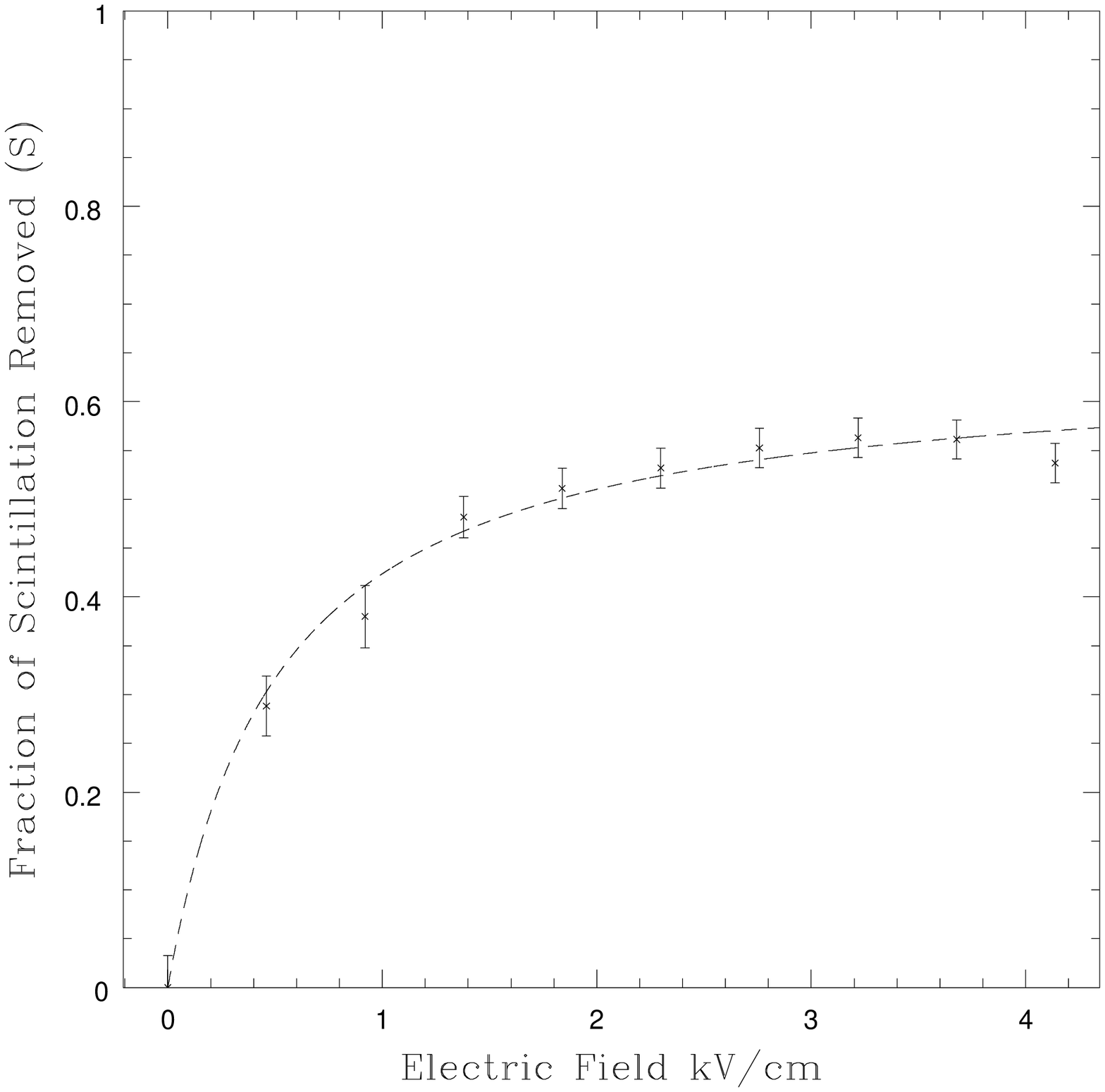}}
\vspace{-0.25cm}
\caption[\small Fraction of the Gamma Ray Scintillation Signal Removed as
a Function of the Electric Field.  ]{\small Fraction of the gamma ray scintillation signal removed as
a function of the electric field.  The dashed line is a line of best
fit (Equation \ref{eqn:co60scintrend} rearranged). \label{fig:gammaareatrend}}}
\end{figure}

For each applied potential difference, pulse height spectra were recorded, both with and without the \nuc{60}{Co} source.  The background spectra (without the \nuc{60}{Co} source) is predominantly from the internal \nuc{214}{Am} souce, see Figure \ref{fig:am241spectrum}.  The background spectra were subtracted from the source spectra to give the \nuc{60}{Co} spectra.  The position of the peak in each \nuc{60}{Co} spectrum was determined.
With the application of an electric field, the position of the peak
moved to lower channels as a result of the scintillation signal area
decreasing.  Figure \ref{fig:gammaareatrend} shows the fraction of the
scintillation signal removed ($S$) as a function of the applied electric
field ($E$).  A straight line fit to the inverse of this quantity
($\frac{1}{S}$), and the inverse of the electric field ($\frac{1}{E}$)
was made, resulting in:
\begin{equation}
\frac{1}{S} = (0.8 \pm 0.1) \frac{1}{E} + (1.56 \pm 0.06).\label{eqn:co60scintrend} 
\end{equation}

\begin{figure}
\centering{ 
\scalebox{0.4}{\includegraphics{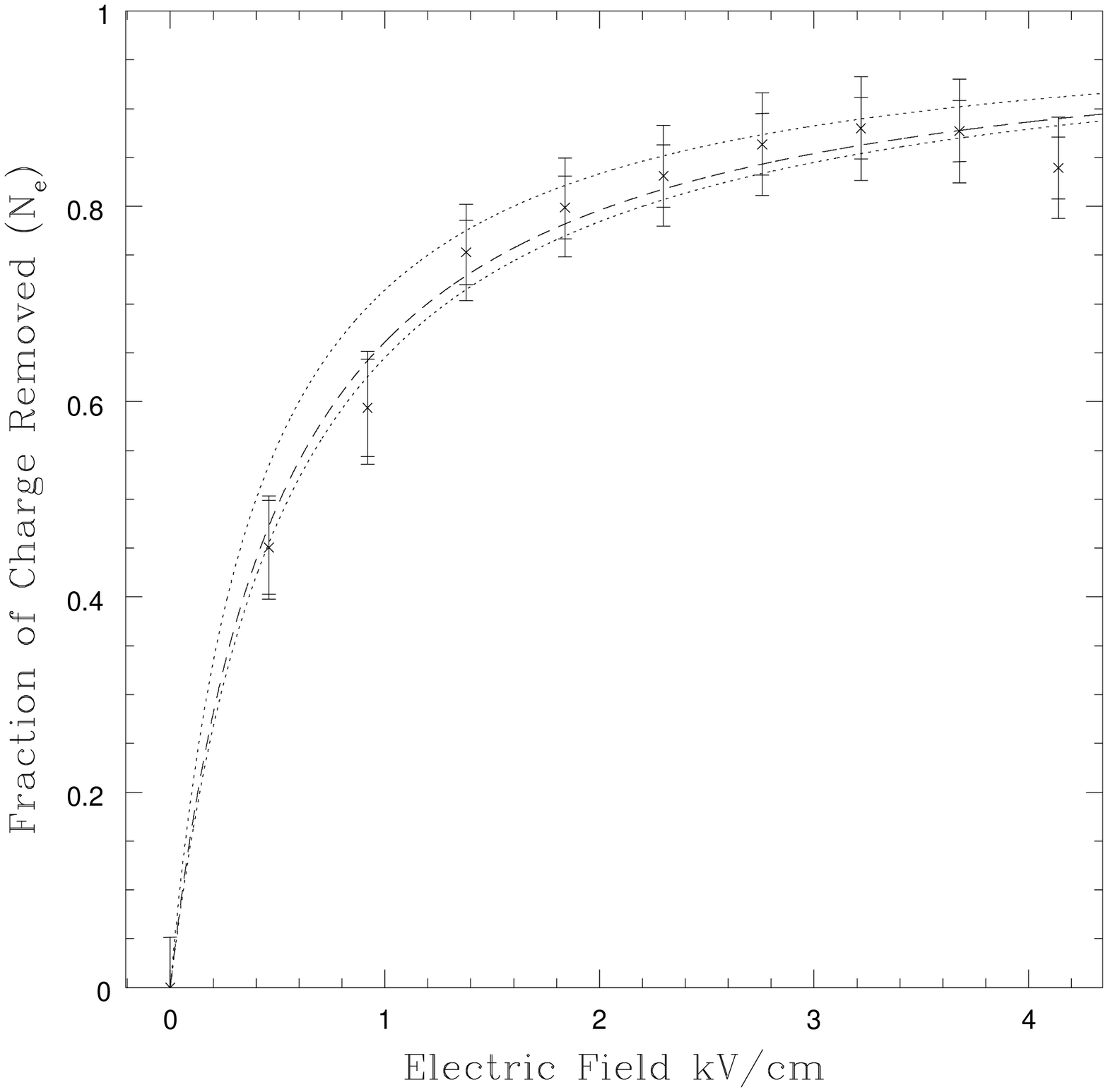}}
\vspace{-0.25cm}
\caption[\small Fraction of Charge extracted as a Function of
Applied Electric Field. ]{\small Fraction of charge extracted as a function of
applied electric field.  The inner error bar represents statistical errors only.  The outer error bars represent total errors including systematic errors.  The dashed line is the line of best fit to
the data (from Equation \ref{eqn:co60scintrend}).  The dotted lines
are from Equation \ref{eqn:davidgeparameterisation}, assuming a gamma
ray energy of 200~keV (lower curve) and 1~MeV (upper curve).  \label{fig:gammanetrend}}}
\end{figure}

For an infinite electric field the percentage of signal removed by the
electric field is equal to the inverse of the intercept, 64~$\pm$~2~\%.  
Work by Kubota, with \nuc{207}{Bi} conversion electrons of energies
0.976 and 1.05~MeV, found a reduction fraction of 0.74~$\pm$~0.02~\cite{Kubota1978} at 12.7~kV/cm, significantly greater than found
in this work.  This suggests a variation in the mean electron
creation energy W$_{e}$ with recoiling electron energy.  

The corresponding fraction of charge
extracted as a function of the electric field is shown in
Figure~\ref{fig:gammanetrend}.  Also shown are two trends from
Equation~\ref{eqn:co60scintrend}.  The lower dotted curve is for a
gamma ray energy of 200~keV, and the upper curve is for an energy of
1~MeV.  The line of best fit to the data lies at slightly higher
energies than the 200~keV trend.  The energies deposited by the large
angle scatters peak at 212~keV, resulting in a faster
increase in charge removal with increasing electric field strength.
For energy deposits greater than 1~MeV, the trends are far less dependent
on the energy deposit and all gamma rays behave in a similar manner;
this is the maximal charge removal trend.

\section{Scintillation Traces}

\begin{figure}
\centering{ 
\scalebox{0.45}{\includegraphics{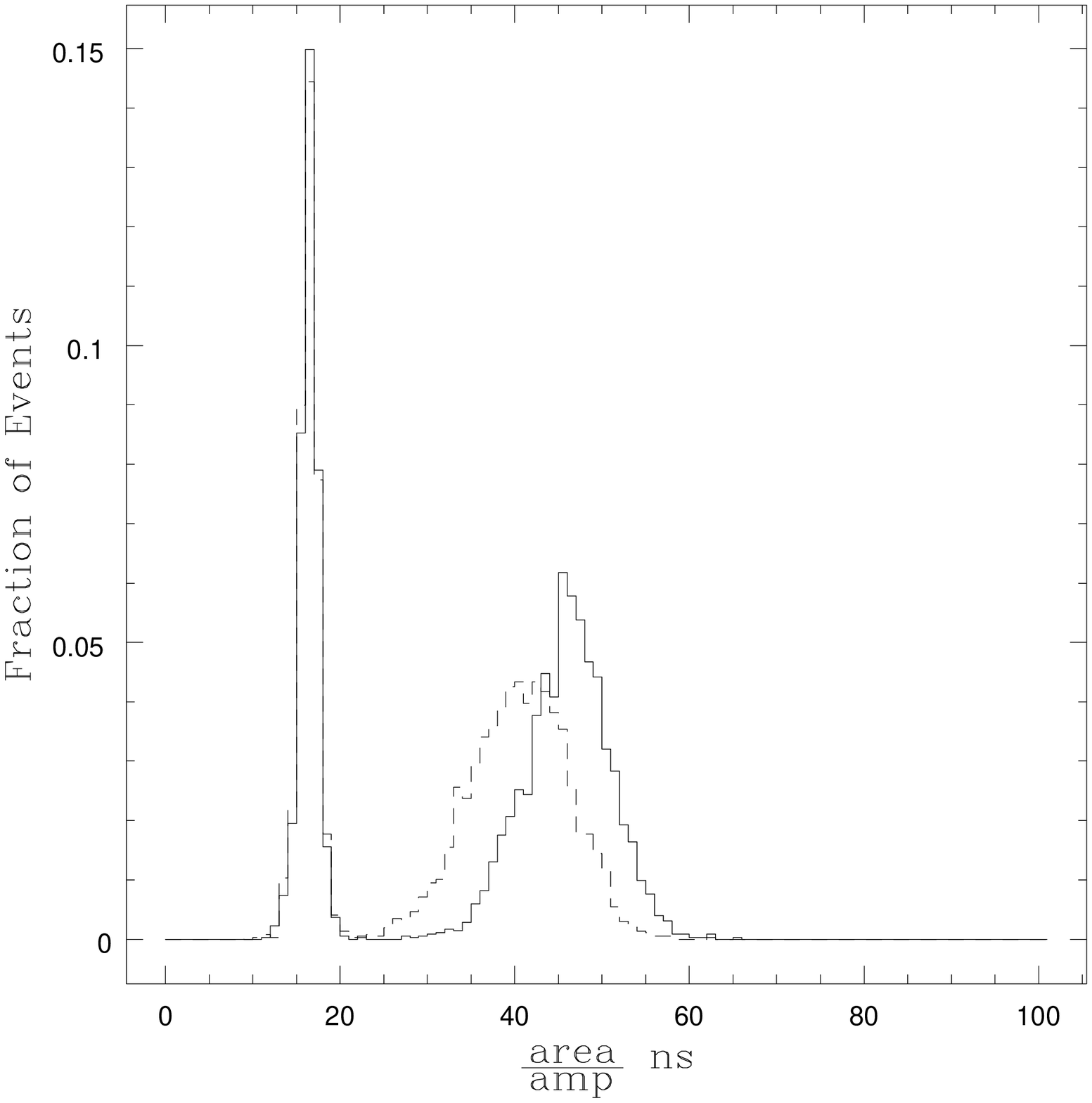}}
\caption[\small  Histograms of the Ratio of Pulse Area to
Amplitude.]{\small Histograms of the ratio of pulse area to amplitude.
 The solid line represents the condition of no applied electric field,
and the dashed line is for an electric field of 0.05~kV/cm.  The alpha
signals have a ratio of $\sim$~15~ns, and display no discernable
change with electric field.  The gamma ray signals have area to
amplitude ratios greater than $\sim$~22~ns, which reduce with increasing
electric field. \label{fig:alpgamhist0and1field}}
}

\centering{ 
\scalebox{0.45}{\includegraphics{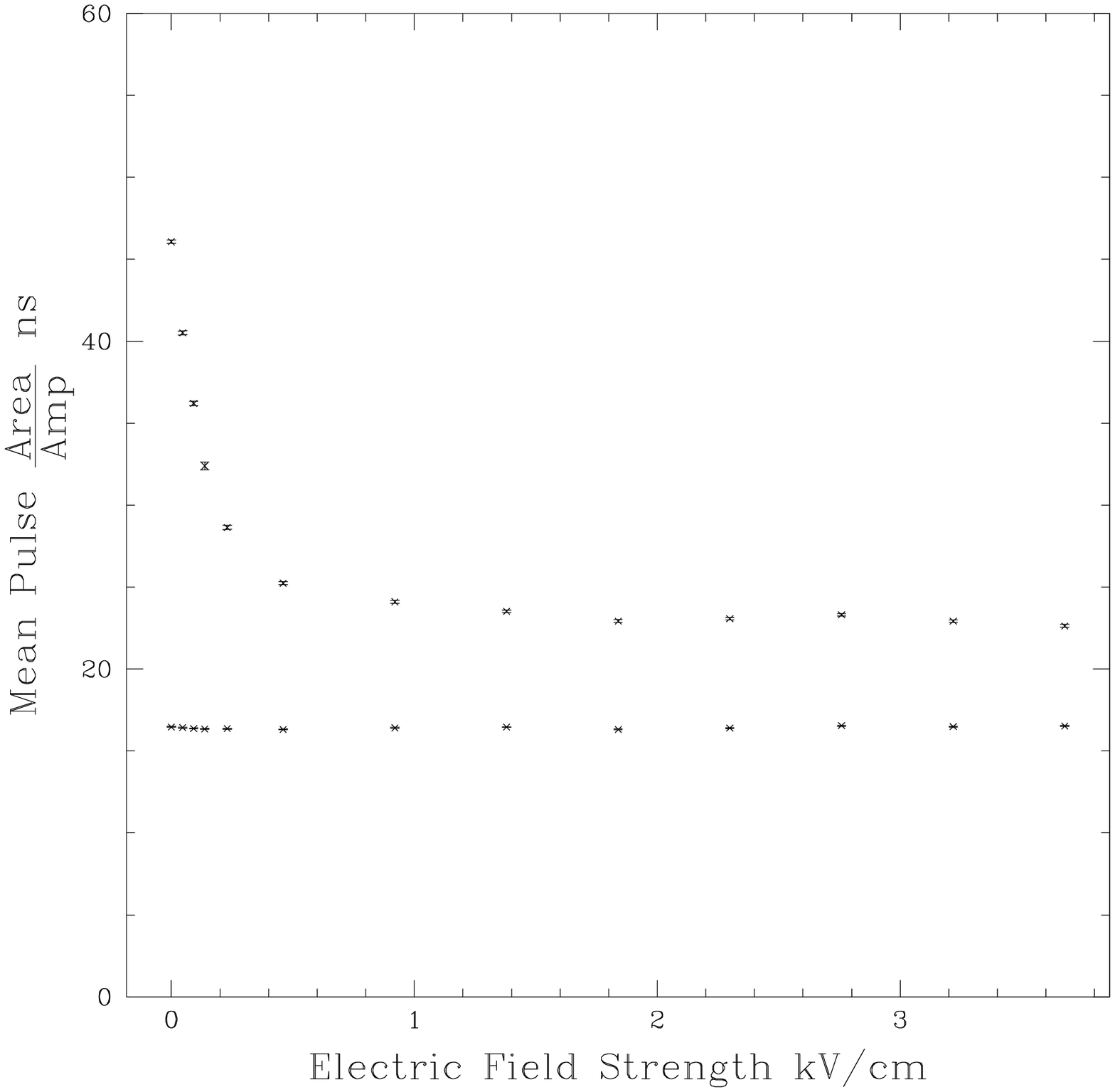}}
\vspace{-0.25cm}
\caption[\small Variation of Pulse Area to Amplitude Ratio Parameter
with Applied Electric Field. ]{\small Variation of pulse area to
amplitude ratio with applied electric field for source alphas and
gammas from an external \nuc{60}{Co} source.  The alphas (lower points)
show no change with increasing electric field.  The gamma rays (upper
points) decrease rapidly with increasing electric field from 46.1~$\pm$~0.1~ns with no electric field to 22.6~$\pm$~0.1~ns at 3.7~kV/cm. \label{fig:alphgammatrend}}}
\end{figure}

Scintillation traces were recorded at each applied electric field
strength.  The photomultiplier voltage output was digitised by a
Lecroy 7200 oscilloscope of 1 Gs/s at 500 MHz bandwidth, and recorded
by computer with Labview DAQ software.  A leading edge trigger
provided the lower amplitude threshold of 0.1~V.  The upper amplitude
threshold was set to 0.9~V through DAQ software selection.  As alpha
and gamma ray signals have different pulse shapes the amplitude
threshold settings select different energy ranges for alphas than for
gamma rays.  For gamma rays, the recorded event energies were between
approximately 100~keV and 1~MeV with no applied electric field.  Alpha
events from the internal $^{241}$Am source have visible energies of
between 50~keV and 400~keV.  These alpha events are the lower energy
tail of the alpha source energy distribution and the threshold
settings used do not include the peak of the spectrum (see Figure
\ref{fig:am241spectrum}).  The recorded traces were analysed through
software, returning maximum amplitudes (pulse amplitudes in V) and
pulse areas (in nVs).  A characteristic time parameter, known as the
time-constant, was defined as the ratio of the pulse area to amplitude
(units of ns).  Figure~\ref{fig:alpgamhist0and1field} shows this
quantity for the condition of no applied electric field (solid) and
0.05~kV/cm (dashed).

The data are clearly divided into alpha and gamma ray populations, based upon
the measured area to amplitude ratios.  Events with time-constants less than
22~ns were deemed alphas, events with ratios greater than this were
identified as gamma rays.  Figure~\ref{fig:alpgamhist0and1field} illustrates
the effect of the electric field on the gamma ray scintillation pulse
shape, reducing the time-constant.  At each electric field value
the mean time-constant was determined for alpha particles and
gamma rays, see Figure~\ref{fig:alphgammatrend}.

\begin{figure}[ht]
\noindent
\unitlength1cm
\begin{minipage}[t]{7cm}
\begin{picture}(7,9)
\centering{
\hbox{\psfig{file=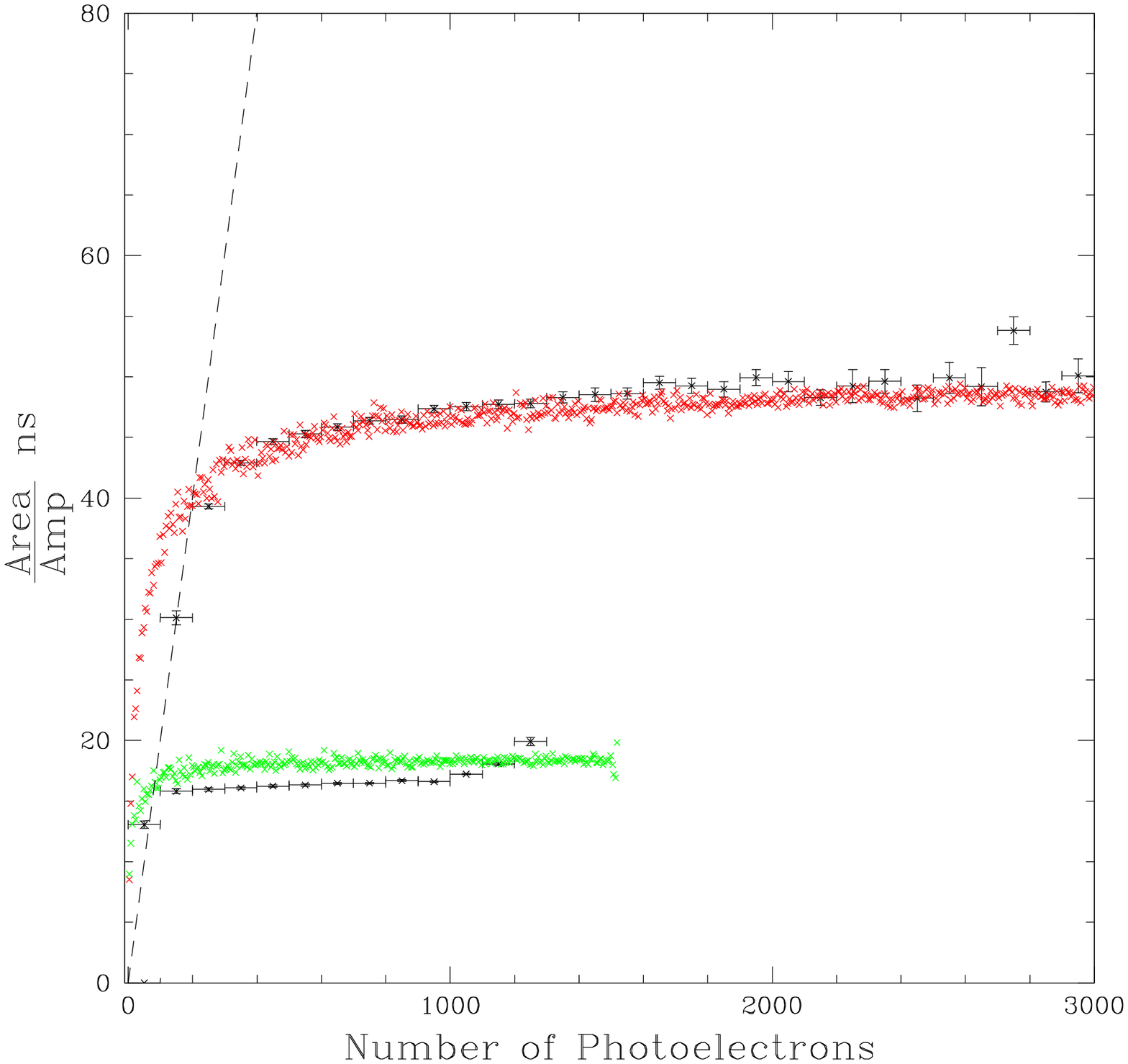,height=9cm,width=7cm}\hspace{1.0cm}}}
\end{picture}
\par
\end{minipage}
\begin{minipage}[t]{7cm}
\begin{picture}(7,9)
\centering{
\hbox{\psfig{file=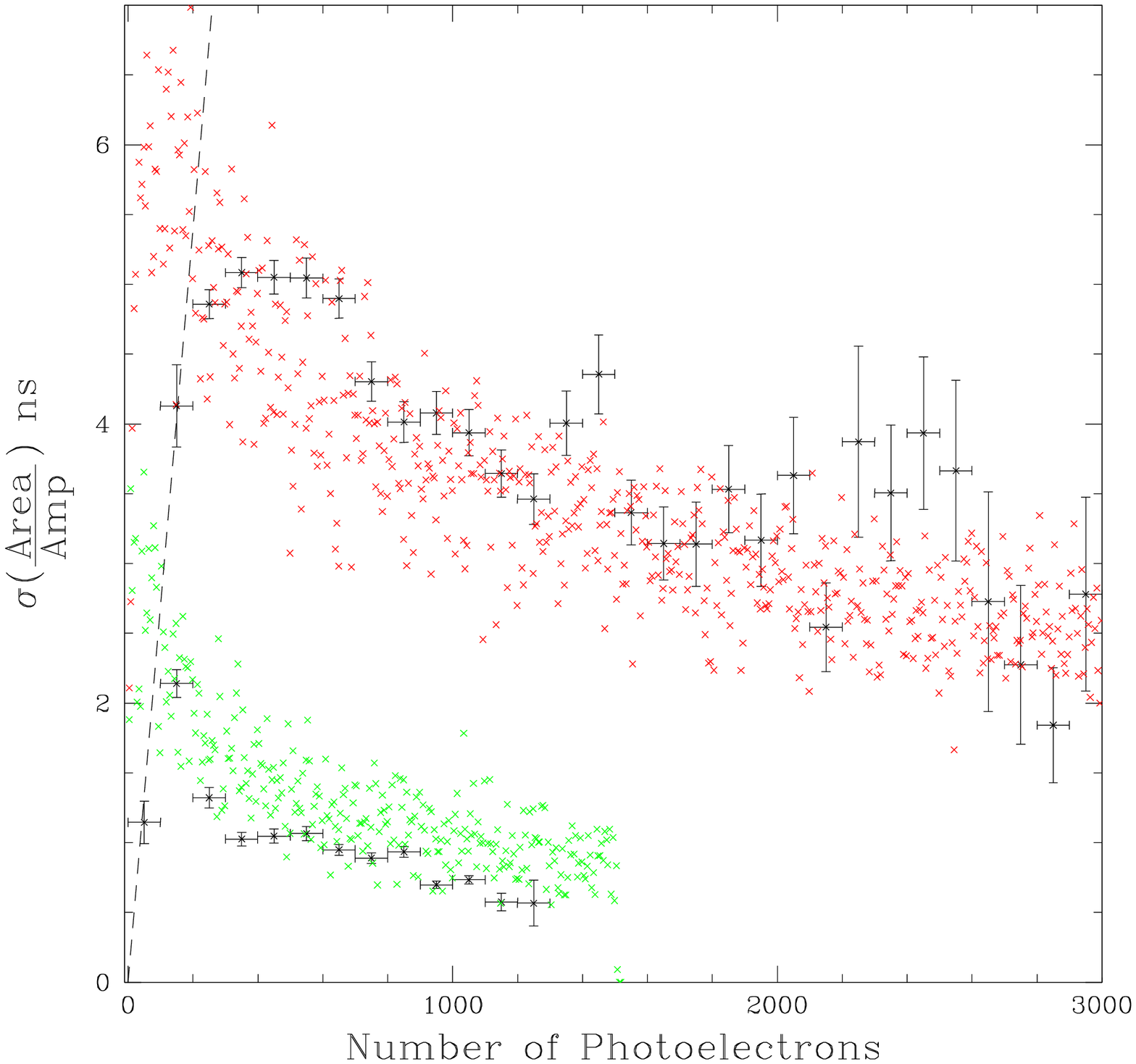,height=9cm,width=7cm}\hspace{1.0cm}}}
\end{picture}
\par
\end{minipage}
\hfill
\vspace{-1cm}
\caption{\small Comparison between the simulated dataset (denoted by
  crosses) and binned experimental data (points with error bars).  On
  the left the mean area to amplitude ratio is shown.  On the right
  the standard deviation of the area to amplitude ratio is plotted.
  The dashed lines represent the effect of lower amplitude threshold on the
  experimental dataset, pulses with amplitudes smaller than this threshold
  are not recorded.
 \label{fig:comparesim}}
\end{figure}

A Monte Carlo simulation was used to investigate the behaviour of
measured time-constant with number of photoelectrons detected.  A
parent exponential time-constant was chosen to mimic the behaviour of gamma rays
under no electric field (47~ns), for alphas a shorter time-constant was
chosen (15~ns).  Exponential random deviates were generated,
simulating the emission time of scintillation photons.  All emitted photons are assumed to strike the photocathode of the photomultiplier tube, and an output voltage trace is produced.  The photomultiplier simulation included the quantum efficiency of the
tube, single photoelectron temporal profile, and jitter.  The dynode
resistor chain, and applied voltage were also implemented to give a
realistic single photoelectron spectrum.  The simulated voltage traces
were analysed, returning pulse areas and amplitudes.  The data were
binned by pulse area and for each bin the mean and standard deviation
of the pulse area to amplitude ratios was found.  The simulated data
were then compared to experimental data, as shown in
Figure~\ref{fig:comparesim}.  The simulations can be made to match the
experimental data well by adjusting the the parent exponential
time-constant.  In this figure the simulations are slightly offset from
the experimental data.  Simulations were also made to represent gamma
rays under various electric fields by changing the parent
time-constant.

\begin{figure}
\centering{ 
\scalebox{0.5}{\includegraphics{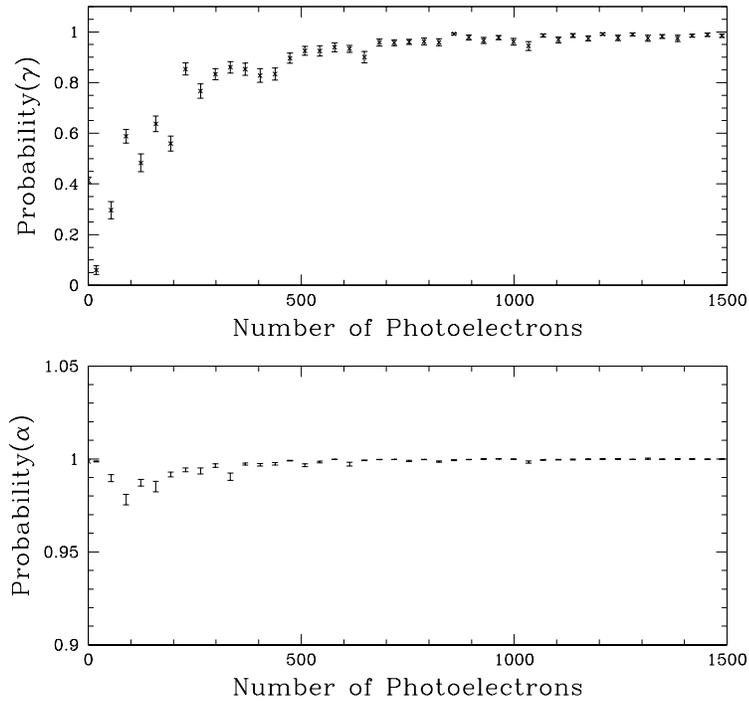}}
\vspace{-0.25cm}
\caption[\small  ]{\small The probability of correctly identifying
  populations using a 22 ns time-constant cut on simulated datasets.  \label{fig:dplot1}}}
\end{figure}

The affect of the 22 ns time-constant cut applied to the real datasets was
explored by applying the same cut on the simulated dataset.  The
probability of correctly identifying alphas (with time-constants less
than 22 ns) and gamma rays (with time-constants greater than 22 ns)
can be assessed.  Figure \ref{fig:dplot1} shows the identification
probabilities determined by simulation as a function of number of
photoelectrons, for the case of greatest applied electric field.  The
alpha particles are well identified with this cut for all
photoelectrons.  The alpha identification probability falls with
decreasing number of photoelectrons as the standard deviation of the
time-constant distribution increases (see right-hand plot, Figure~\ref{fig:comparesim}).  Below a hundred photoelectrons, the
probability increases as the mean of the time-constant distribution
falls (see left-hand plot, Figure~\ref{fig:comparesim}) overriding the
effect of the increasing standard deviation.  For gamma rays, the
identification probability falls with number of
photoelectrons, due to both the mean and standard deviation of the
time-constant distribution falling and rising respectively.  For a gamma ray signal comprising 1000 photoelectrons
the probability of correctly identifying the signal is 0.96 $\pm$ 0.01.  Equivalently for an alpha signal the probability is
greater than 0.999.  For the experimental datasets (with large numbers
of photoelectrons) this time-constant cut was good and should not have
biassed the results, even under high electric field strengths.  However, using
this cut on signals with low numbers of photoelectrons would have
resulted in misidentifying many gamma rays as alpha particles.  To
achieve the most success in separating gamma rays from alpha
particles, the time-constant cut should vary with number of
photoelectrons and be set according to one of three conditions; to
either accept fewer alphas (and reject gammas with high efficiency),
to accept fewer gammas (and reject alphas with high efficiency), or to
some intermediary state in which the majority of one population are
accepted but with a significant contamination of the other population.  As
an example, with 100 photoelectons and a 22 ns cut the probability of
identifying a gamma ray has a low efficiency ($\sim$ 0.5) but the
probability of an alpha particle contaminating the sample is unlikely (less than 0.02).  In this case, the gamma rays are
identified with a low efficiency but alpha particles are highly
discriminated. 

\section{Conclusion}
\label{Conclusion} 

Pulse height spectra produced by an external \nuc{60}{Co} source and
the internal \nuc{241}{Am} source were compared to GEANT~4~\cite{Geant42003}
simulations.  The features of both spectra are well reproduced.  
For the majority of
this sensitive volume the yield is estimated to be 1.5~photoelectrons
per keV (under the condition of no applied electric field).

The variation of scintillation pulse area from \nuc{60}{Co} gamma rays
was investigated as a function of electric field, using the large-angle scatter
peak.  
The scintillation pulse areas decrease as the applied electric
field increases. 
For a 212 keV gamma
ray, 64~$\pm$~2~\% of the scintillation is found to be produced via
recombination.  


The temporal profile of the scintillation signal was investigated for
incident alpha particles and gamma rays with increasing electric
field.  No significant change with electric field was observed for
alpha particles.  Gamma ray signals, however, become faster with low electric field strengths from
46.1~$\pm$~0.1 ns with no electric field to 25.2~$\pm$~0.1~ns at
0.5~kV/cm.  Increasing the electric field strength further reduces the
time-constant to 22.6~$\pm$~0.1~ns at 3.7~kV/cm.  The behaviour of the derived
time-constant with numbers of photoelectrons detected was investigated
with simulation and compared to experimental data.  The behaviour of
measured time-constants from alphas and gammas (at all fields) can be
reproduced using only one parameter, the parent time-constant.

From this work, the scintillation signal (both in size and shape) produced by
any energy deposit from an alpha particle or gamma ray in liquid xenon
can be predicted.  The number of electrons extracted from a gamma ray
interaction under any strength electric field can also be predicted,
aiding simulations of the response of two-phase xenon detectors to
gamma ray interactions.

\end{document}